\documentclass[12pt,a4paper]{article}

\usepackage[greek,american]{babel}
\usepackage{amssymb}
\usepackage{amsfonts}
\usepackage{amsmath}
\usepackage{relsize}
\usepackage{subfigure}
\usepackage{float}
\usepackage{mathtools}
\usepackage{dcolumn}% Align table columns on decimal point
\usepackage{bm}
\usepackage{subfloat}
\usepackage{xcolor}
\usepackage{graphicx}
\usepackage{enumitem}
\usepackage{algorithm}
\usepackage{algpseudocode}
\usepackage{pifont}

\usepackage{bbm}

\newcommand{\iid}{\stackrel{{\rm i.i.d.}}\sim}

\def\G{\mathbb G}
\def\R{\mathbb R}

\def\R{\mathbb R}
\def\E{\mathbb E}
\def\X{\mathbb X}

\def\X{\mathbb X}

\def\vt{\vartheta}
\def\l{\lambda}
\def\a{\alpha}
\def\b{\beta}

\def\d{\delta}
\def\t{\theta}

\def\s{\sigma}

\begin{document}

	\begin{center}
		{{{\Large\sf\bf A Bayesian nonparametric approach to the approximation of the global stable manifold}}}\\
		
		\vspace{0.5cm}
		{\large\sf Spyridon J. Hatjispyros
			\footnote{Corresponding author. Tel.:+30 22730 82326\\
				\indent E-mail address: schatz@aegean.gr}, Konstantinos Kaloudis	}
		
		\vspace{0.2cm}
	\end{center}
	
	\centerline{\sf Department of Statistics and Actuarial--Financial Mathematics, 
		University of the Aegean}
	\centerline{\sf Karlovassi, Samos, GR-832 00, Greece.} 

\begin{abstract}
We propose a Bayesian nonparametric model based on Markov Chain Monte Carlo
(MCMC) methods for unveiling the structure of the invariant global stable manifold
from observed time-series data. The underlying unknown dynamical process is possibly
contaminated by additive noise. We introduce the Stable Manifold Geometric Stick Breaking
Reconstruction (SM-GSBR) model with which we reconstruct the unknown dynamic equations and 
in parallel we estimate the global structure of the perturbed stable manifold. Our method 
works for noninvertible maps without modifications. The stable manifold estimation procedure
is demonstrated specifically in the case of polynomial maps. Simulations based on synthetic time series are presented.

\vspace{0.1in}\noindent 
\end{abstract}

%\begin{quotation}
%We assume that the underlying stochastic model is amenable 
%to estimation via the Geometric Stick Breaking Reconstruction (GSBR) method introduced in Ref. [\onlinecite{merkatas2017bayesian}] i.e. in the form of a deterministic part and an additive 
%dynamical noise component which is (perhaps) non-Gaussian. We introduce the SM-GSBR model, 
%a Gibbs sampler scheme, with which, in parallel with the reconstruction process of the unknown 
%deterministic part and noise component, we sample from the supports of remote in the past out-of-sample observation variables. Such variables have full conditional densities, depending exclusively on observations that are ahead in time, thus making their support to diffuse along the direction of the local stable manifold contained on a neighborhood of the unknown true initial condition of the 
%observed time-series. 
%We show that the union of such supports, conditional on proper sliding 
%window subsets of the time-series data, contain information for the global stable manifold of the underlying
%system. 
%\end{quotation}

\section{Introduction}
\label{sec:Intro}
In the recent literature, there has been a growing interest for the construction and analysis of models that include nonlinear deterministic components and are influenced by a variety of noise processes. There exist two main distinct fields of research associated with random perturbations in dynamical systems. The first approach regards the analysis of Random Dynamical Systems, by defining direct stochastic generalizations of the central notions appearing in the theory of Dynamical Systems, such as random attractors \cite{crauel1997random} and random invariant manifolds \cite{mohammed1999stable}. The second approach, which we will follow in this paper,  relates to the behavior of the perturbed system to its zero-noise limit, that is having as a reference system the underlying nonlinear deterministic part.

Invariant manifolds are essential to the theory of dynamical systems, as the behavior of any dynamical system is related to the underlying geometrical structure of the state space, more specifically the organization of the invariant stable and unstable subspaces. There exist several methods in the literature suitable for the approximation of invariant manifolds of deterministic and random dynamical systems.

The need for a development of generic methods suitably approximating the invariant manifolds of nonlinear dynamical systems is highlighted by the fact that in general it is not possible to describe the invariant set of points via closed-form analytic expressions. A special case of a power series expansion approximation for the stable and unstable manifolds of a H\'{e}non map saddle fixed point is presented in Ref. \cite{franceschini1981stable}. The most widely used method for the approximation of the global stable manifold of any invertible map is presented in Ref. \cite{you1991calculating}, while the authors also propose a method suitable for noninvertible maps in \cite{kostelich1996plotting}. Another method for the approximation of the global stable manifold of invertible maps is introduced in \cite{krauskopf1998growing} and is extended to the Search-Circle algorithm \cite{england2004computing} for the noninvertible case. A modified version of the Search-Circle, which is faster and is based on the same concept is presented in \cite{li2012new}. For the numerical computation of higher-dimensional invariant manifolds, we refer to \cite{gonzalez2017high, krauskopf1999two,guckenheimer2004fast}. For a review of numerical methods for two-dimensional manifolds we refer to \cite{krauskopf2005survey}. As a restriction to the above methods, can be considered the requirement of the complete knowledge of the dynamical system. A method suitable for the approximation of the invariant manifolds given only experimental data, and when there is no available model, is proposed in Ref. \cite{triandaf2003approximating}. Moreover, the ``parameterization method'' originally introduced by Cabr\'{e} et. al \cite{cabre2003parameterization, cabre2003parameterizationb}, forms the basis of a series of results relevant to the establishment of the existence and the computation \cite{van2016computing} of invariant manifolds.

In this work, we will use a Bayesian nonparametric approach to reconstruct the unknown random dynamical system and simultaneously jointly estimate variables of past unobserved-observations
based solely on time series data.
More specifically, we propose an extension of the GSBR sampler introduced in Ref. \cite{merkatas2017bayesian}, in order to provide a MCMC based stochastic approximation of the global stable manifold based on observed noisy time-series. We introduce the 
Stable Manifold Geometric Stick Breaking model, from now on referred to as the SM-GSBR model.
Our intention is to compute the support of the marginal posterior densities of certain sets of past unobserved-observations by performing predictions in reversed time. We emphasize on the diffusive support of marginal posterior densities of far-off past random variables, modeling initial conditions. We demonstrate that such supports, contain part of the associated noisy stable manifold. Finally, when the SM-GSBR sampler is applied on proper sliding window time-series, 
the global noisy stable manifold reveals itself
naturally as the union of the supports of the posterior marginal distributions of random variables describing remote in the past initial conditions.

The unique feature of the random variable which is modeling a starting point for an observed time-series, is that it is the only variable with a full conditional density, depending exclusively on observations that are ahead in time. All other variables, have full conditional densities that depend on both future and past observations. This apparent lack of information regarding past observations, makes the support of the associated full conditional to diffuse along the direction of the local stable manifold contained on a neighborhood of the true initial condition. We will see that when we perform predictions in reversed time, the fluctuations of the far-off past variables describing initial conditions is increased, thus spreading the sampled values along the stable direction.

Likewise, conditionally on proper subsets of the observed noisy time-series, we can
approximate parts of the global unstable noisy manifold, by means of the diffusive 
supports of the random variables describing the ending points of forward out-of-sample 
future unobserved-observations.    

We will demonstrate the efficiency of the proposed method using different types of polynomial maps, which are of particular interest not only because of their rich dynamical behavior, but also due to their ability to approximate more complicated maps, as finite degree Taylor approximations of non-polynomial nonlinearities. Furthermore, in essence all invertible polynomial maps, i.e. any nontrivial polynomial diffeomorphism of $\R^2$ with constant Jacobian, are conjugate to compositions of generalized H\'{e}non maps \cite{dullin2000generalized}.

The basic advantage of the proposed method is its ability to provide an adequate stochastic approximation of the stable manifold, under a data driven method. In particular, no knowledge regarding the parameters of the system is required, not even the location of the saddle fixed point, just a general functional representation of the deterministic part. Namely, the procedures of the system identification and of the stochastic approximation are performed 
{\it in parallel}, in a similar fashion as the Dynamic Noise Reduction Replicator model described in Ref. \cite{kaloudis2018bayesian}. Our method is parsimonious, due to its flexibility induced by the general functional form of the deterministic part, and the application of a GSB mixture process prior over the additive stochastic component. Our 
method is applicable under small data sets which are possibly corrupted by (perhaps) non-Gaussian noise. Moreover, an important feature of the SM-GSBR model is its wide applicability. In fact we will demonstrate the applicability of the SM-GSBR algorithm equally to the case of invertible and the case of noninvertible maps.

This work relates to utilizing a Bayesian nonparametric framework, 
in order to approximate dynamical invariants, based entirely on observed time 
series data. As we have seen in a series of previous works, one can in principle: 
\begin{enumerate}
	\item 
	 Approximate the quasi-invariant measure, of the underlying random dynamic system, 
	 thus implying the existence of a prediction barrier \cite{merkatas2017bayesian}.
	
	\item 
	Calculate the positions of primary homoclinic tangencies, 
	revealing the non-hyperbolic nature of the underlying perturbed deterministic dynamic 
	system \cite{kaloudis2018bayesian}.
  
  \item
  Utilize borrowing of strength relationships among the dynamical error pairs of multiple
  time-series, such that underrepresented data sets benefit in terms of model estimation
  accuracy \cite{Hatjisp2019}.
\end{enumerate}

The layout of the paper is as follows. In Section \ref{s2}, we discuss some preliminary notions. In Section \ref{s3}, we derive the SM-GSBR model, a Bayesian nonparametric model incorporating a GSB 
mixture process prior, for making predictions in reversed time, suitable for the stochastic approximation of the global stable manifold from observed time-series data. In Section \ref{s4}, we resort to simulation. We apply the proposed SM-GSBR model on deterministic and dynamically perturbed versions 
of the classical H\'enon map, the Dual-H\'enon map and a noninvertible map with a H\'enon-like strange
attractor. Finally, in Section \ref{s5} are the concluding remarks.

%------------------------------------------------------------------------------------------------

\section{Preliminaries}\label{s2}

In what follows, we will consider a planar diffeomorphism $g$ of $\R^2$ to itself.
The global dynamical behavior of the map $g$ is determined by dynamically invariant objects, such as the stable and the unstable manifolds. Let $y$ be a saddle fixed point of the map $g$, i.e. the Jacobian matrix $D_y\, g$ has eigenvalues $\l_{\rm s}$ and $\l_{\rm u}$ such that $|\l_{\rm s}|< 1$ and $|\l_{\rm u}|>1$. The global stable manifold \cite{gilmore2002topology, krauskopf1998growing} $W^{s}(y)$ of $y$, is the set of points 
whose orbit tend to $y$ in forward time, namely
$$
W^{s}(y) = \left\{x \in\R^2: g^n(x)\rightarrow y, n \rightarrow \infty  \right\}.
$$
Similarly, the corresponding unstable manifold is defined as the set of points whose orbits tend to $y$ in reversed time, that is
$$
W^{u}(y) = \left\{x \in \mathbb{R}^2: g^{-n}(x)\rightarrow y, n \rightarrow \infty  \right\}.
$$
Since $g$ is invertible, the stable manifold of $g$ is the unstable manifold of $g^{-1}$ and vice versa.

Due to the stretching and folding mechanisms, the two manifolds do not only intersect at the fixed point but also in other locations called homoclinic points. If the intersection $p$ between the invariant manifolds is tangential, we have a point of homoclinic tangency (HT).
There are infinitely many HTs, due to the fact that $g(p)$ and $g^{-1}(p)$ are themselves HTs. HTs are important in the study of how noise affects chaotic systems, as they lead to noise amplification. In the symbolic dynamics literature, they have been proposed as a basis for building generating partitions. Furthermore, the existence of HTs is mutually exclusive with the property of hyperbolicity, thus making shadowing impossible. Dynamical perturbations in the vicinity of HTs are driving away the dynamics from the neighborhood of the attractor. 
Nevertheless, due to the folding effect of the nonlinear map, the noisy orbits are mapped onto the neighborhood of the attractor again. 
For a detailed description of the above mechanism see Ref. \cite{jaeger1997homoclinic}.

The H\'enon map\cite{henon1976two}, is defined as
\begin{equation}
\label{henon}
(x,y)\mapsto g(\vt,x,y) = (\theta_1 + \theta_2 x^2 + \theta_3 y, x),
\end{equation}
%with inverse:
%\begin{equation}
%g^{-1}\left(\boldsymbol{\theta},x,y\right) = \left(x, -\dfrac{1}{\theta_3}(\theta_1 + \theta_2 y^2 - x) \right).
%\end{equation}
where $\vt$ denotes the dependence of $g$ to control parameters.
The H\'enon map exhibits the same qualitative dynamical behavior with a wide variety of polynomial maps, namely quadratic maps with a constant Jacobian.

\begin{figure}[hbt!]
	\centering
	\includegraphics[width =0.7\textwidth]{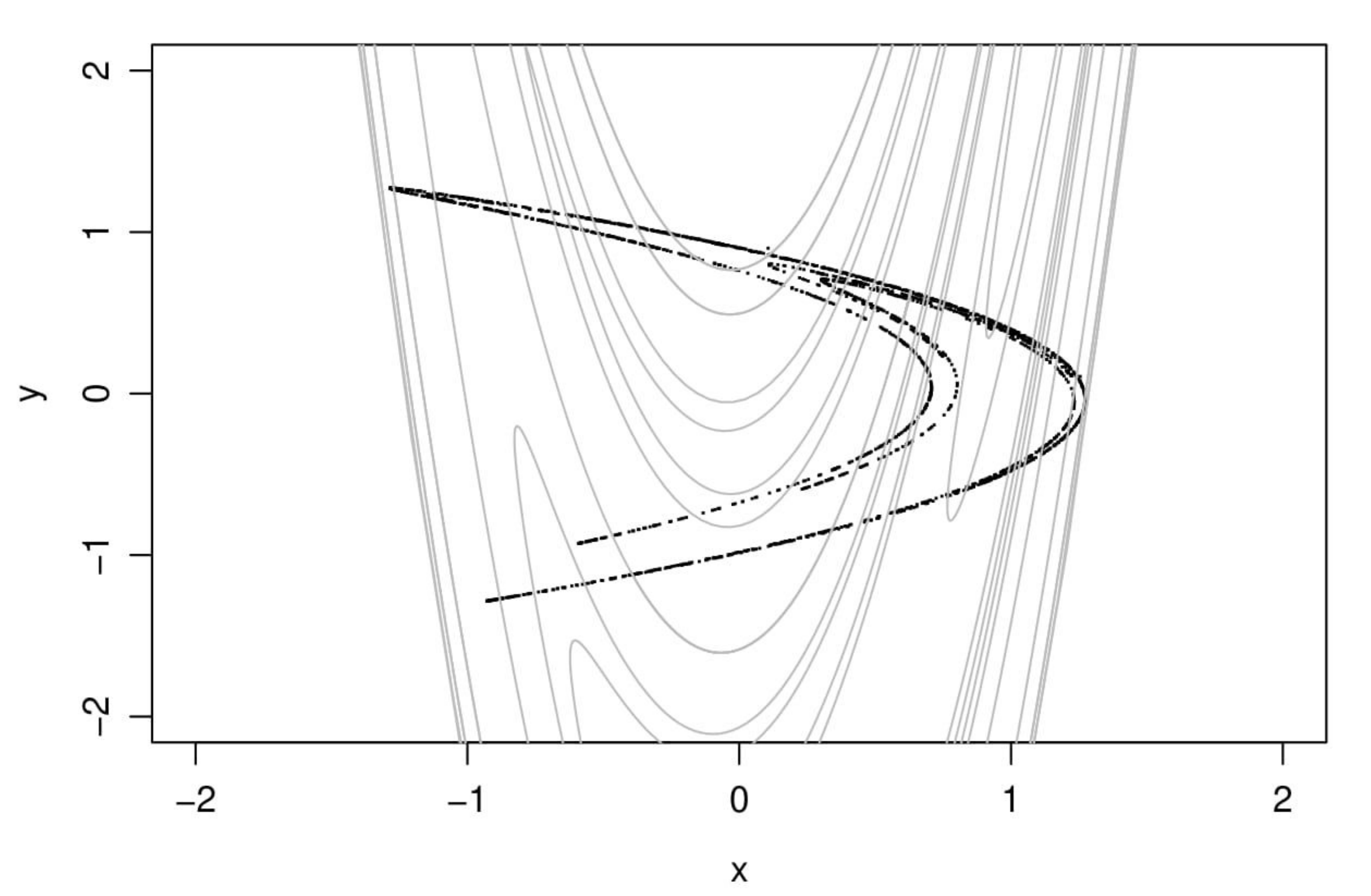}
	\caption{The attractor of the H\'enon map (in black), is superimposed with portions of the
	         global stable manifold (in gray). 
	         \label{fig:henonsm}}
\end{figure}

In Fig. \ref{fig:henonsm} we present portions of the stable manifold of the saddle fixed point $(0.631, 0.631)$, for the classical parameter values $\vt = (1,-1.4,0.3)$. We obtain points of the stable manifold by iterating a set of points in small line segments, aligned along the stable direction of the saddle fixed point \cite{gilmore2002topology}. 
We notice the emergence of homoclinic tangencies, as the map exhibits non-hyperbolic chaotic behavior. The attractor has been conjectured to be the closure of the unstable manifold.

\begin{figure}[hbt!]
	\centering
	\includegraphics[width =0.7\textwidth]{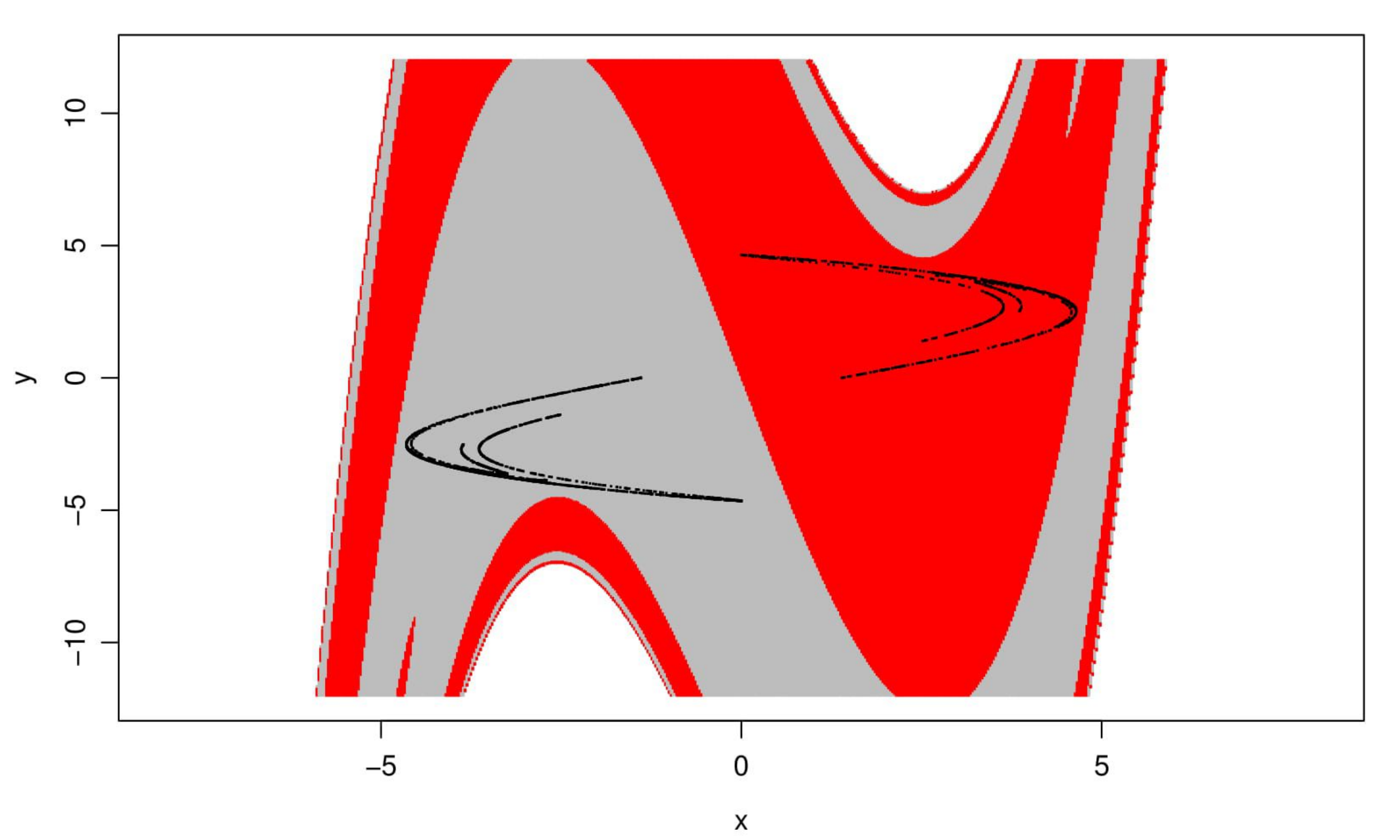}
	\caption{The two coexisting attractors of the Dual-H\'enon map (in black),
	         superimposed with their corresponding basins of attraction 
	         in red and gray.
	         \label{fig:dhba}}
\end{figure}

An illustrating example, regarding the relevance of the stable manifold and the basins of attraction in the case of a multistable behavior, is the Dual-H\'enon map \cite{sprott2015classifying}. The latter map is a special case of the 
Generalized-H\'enon map\cite{dullin2000generalized}, which is defined as
\begin{equation}
\label{dualhenon}
(x,y)\mapsto g(\vt,x,y)=(\t_1+\t_2x+\t_3x^3+\t_4y, x).
\end{equation}
%with inverse:
%\begin{equation}
%g^{-1}\left(\boldsymbol{\theta},x,y\right) = \left(x, -\dfrac{1}{\theta_4}(\theta_1 + \theta_2 y + \theta_3 y^3 - x) \right).
%\end{equation}
When $\vt = (0,2,-0.1,0.3)$, we have the particular case of the Dual-H\'enon map. 
In Fig. \ref{fig:dhba} we present two symmetric, with respect to the origin, isolated 
strange attractors along with their corresponding basins of attraction. The two basins of
attraction are extending to infinity along both directions of the $y$-axis and have a fractal 
boundary \cite{sprott2015classifying}.

\begin{figure}[hbt!]
	\centering
	\includegraphics[width =0.7\textwidth]{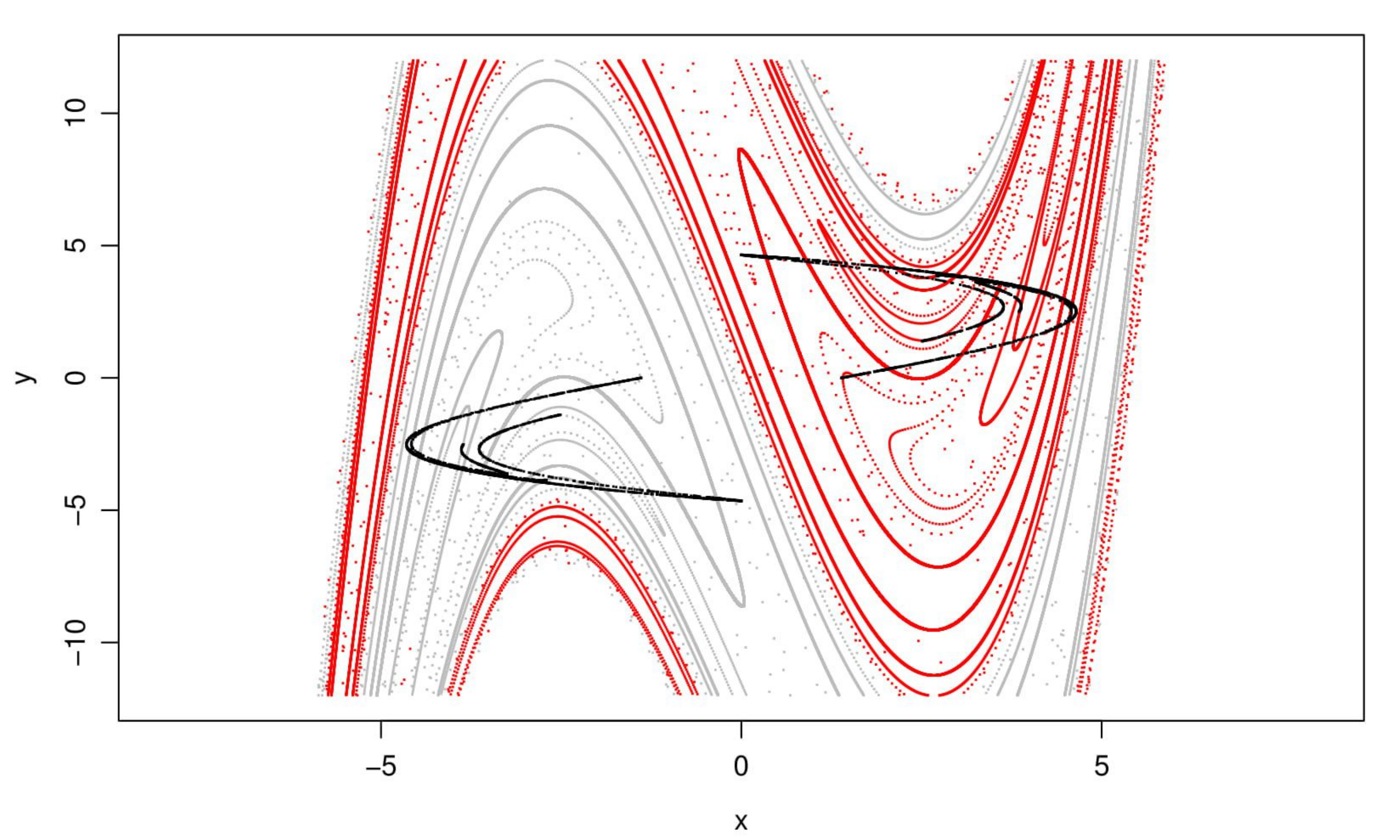}
	\caption{The two coexisting attractors of the Dual-H\'{e}non map, 
	         superimposed with the corresponding
	         global stable manifolds. We have colored red and gray the global stable manifolds 
	         of the coexisting saddle fixed points in the first and second quadrant, respectively.
	         \label{fig:dhsm}}
\end{figure}

In Fig. \ref{fig:dhsm} we plot the pair of strange attractors, along with the two global stable manifolds superimposed. The two associated saddle fixed points are situated in
$(3.605,3.605)$ and $(-3.605,-3.605)$. The interlaced global stable manifolds form the boundaries between the two different basins. Points situated on the stable manifolds,
will eventually converge to the associated saddle fixed points.

%===============================================================================================	

\section{The SM-GSBR model}\label{s3}

Suppose that we have at our disposal, a time-series $x_{1:n}:=(x_1,\ldots,x_n)$ of length $n$, 
which is a realization of the random recurrence relation:
\begin{equation}
\label{realproc3}
X_i={\cal T}(\vt, X_{i-d\,:\,i-1}, e_i)=g(\vt, X_{i-d\,:\,i-1}) + e_i,\,\,i\ge 1,
\end{equation}
where $g:\Theta\times\X^d\to\X$, for some compact subset $\X$ of $\R$. 
The states of the system $(X_i)_{i\ge -d+1}$ and the noise perturbations $(e_i)_{i\ge 1}$,
are all real random variables over some probability space $(\Omega, {\cal F}, {\rm P})$. 	
We denote by $\vt\in\Theta\subseteq\R^m$ any dependence of the deterministic part of the
random map on control parameters. We assume, that the deterministic part $g$ of $\cal T$, 
is nonlinear and continuous in the variable $X_{i-d\,:\,i-1}$.

In addition, we assume that the random variables $e_i$, are independent to each other and independent of the states $X_{i-r}$ for all $r<i+d$. 
We relax the conventional assumption of normality for the additive dynamical errors $e_i$, 
by making the more general assumption, that the $e_i$s are are independent and identically
distributed (i.i.d.), from some symmetric with respect to zero distribution, with unknown 
density $f$ with support over the real line, so that ${\cal T}:\Theta\times\X^d\times\R\to\R$.

We will not take into account observational noise,
so that the time series $x_{1:n}$ is completely specified by the initial
point $x_{-d+1:0}$, the control parameters $\vt$ and the particular realization $e_{1:n}$ 
of the noise process. Also, we note that the observed time-series, can be considered as the one-dimensional measurements of the states of a dynamical system of unknown dimension, which can be embedded in a space of proper dimension, using delay coordinates with a suitably chosen time delay.  

We propose a Bayesian nonparametric model, the SM-GSBR model, in order to jointly estimate a fixed number of consecutive past observations, by performing prediction in reversed time. 
Given the time-series $x_{1:n}$ and a backward prediction horizon $T$, we aim to estimate
the {\it support} of the $T$-shifted to the past initial variable 
$X_{-T-d+1:-T}$, associated with the sequence of variables $X_{-T+1:n}$.

%For example, letting $d=2$ and $T=3$ and given $x^n$, our goal will be to estimate jointly, the three past observations $\tilde{X}_{-3:3}=(X_{-2},X_{-1},X_{0})$ and the support of the initial condition $\tilde{X}_{-5:2}=(X_{-4},X_{-3})$ corresponding to the augmented sequence of variables $\tilde{X}_{-3:n+3}=(X_{-2},X_{-1},X_{0},X_1,\ldots,X_n)$.

%Subsequently, we extend the GSBR model\cite{merkatas2017bayesian}, such that we will be able to  estimate jointly the posterior the control parameters $(\vt|x^n)$, the (perhaps) non-Gaussian density $(f\,|\,x^n)$ of the dynamical noise process, the vector of past observations $(\tilde{X}_{-T:T}\,|\,x^n)$ together with the corresponding initial variable $(\tilde{X}_{-T-d:d}\,|\,x^n)$.

\subsection{The infinite random mixture prior}

We will model a-priori the errors in recurrence (\ref{realproc3}) as a random infinite mixture
of zero-mean normal kernels with mixing measure the GSB random measure $\G$, as
\begin{equation}
\label{randmixt1}
e_i\iid f(x\,|\,\G) = \int_{\tau>0}{\cal N}(x\,|\,0,\tau^{-1})\,\G(d\tau),
\end{equation} 
where $\tau$ is the precision (inverse variance).

The random measure $\G$ is closely linked to the Dirichlet random measure 
\cite{ferguson1973bayesian, sethuraman1994constructive}
$$
\G_{\rm DP}=\sum_{k=1}^\infty w_k\d_{\tau_k}\sim{\cal DP}(c,G_0).
$$
The Dirac measures $\d_{\tau_k}$ are concentrated on the random precisions $\tau_k$, which 
are drawn independently from the mean parametric distribution $G_0$, which is the prior guess
of $\G_{\rm DP}$ i.e. $\E(\G_{\rm DP}(A))=G_0(A)$, for measurable subsets $A$ of $\R_+$.
The weights $w_k$ satisfy $\sum_{k\ge 1}w_k=1$ a.s. and are
stick-breaking, that is $w_1=v_1$ and $w_k=v_k\prod_{l<k}(1-v_l)$, and random 
because $v_k\iid{\cal B}e(1,c)$, a beta density with shapes 1 and $c$. 
The GSB random measure $\G$ is defined by removing 
a hierarchy from the random measure $\G_{\rm DP}$\cite{Fuentes} as
$$
\G=\sum_{k=1}^\infty\E(w_k)\,\d_{\tau_k}\sim{\cal GSB}(\a,\b,G_0).
$$
Then, letting $\l:=\E(v_k)=(1+c)^{-1}$, it is that $\E(w_k)=\l\,(1-\l)^{k-1}$
(the geometric probability of the GSB prior measure), and $\E(\G(A))=G_0(A)$. 
Finally, we randomize the probability-weights of $\G$ by letting $\l\sim{\cal B}e(\a,\b)$.
In Ref.\cite{merkatas2017bayesian}
it is shown that the $\G$-based Bayesian nonparametric framework for dynamical system estimation
is efficient, faster and less complicated
when compared to Bayesian nonparametric modeling via the Dirichlet process.

Taking into account the previous discussion, letting 
$\tau_{1:\infty}=(\tau_j)_{j\geq 1}$ and the fact that the map 
$\G\mapsto (\l,\tau_{1:\infty})$
is one-to-one, the random mixture in (\ref{randmixt1}) can be represented as
\begin{equation}
\label{randmixt2}
e_i\iid f(x\,|\,\l,\tau_{1:\infty}) = \l\sum_{k=1}^{\infty}(1-\l)^{k-1}
{\cal N}(x\,|\,0,\tau_j^{-1}).
\end{equation}

Now, the posterior density $\pi(x_{-T-d+1:-T}\,|\,x_{1:n})$ of the $T$-shifted to the past 
initial condition variable $X_{-T-d+1:-T}$, given the observed time-series $x_{1:n}$,
is a marginal density of the posterior
\begin{equation}
\label{posterior1}
\Pi(\l,\tau_{1:\infty},\vt,\,x_{-T-d+1:-T},\,x_{-T+1:0}\,|\,x_{1:n}).
\end{equation}

Nevertheless, the prior mixture density over the additive errors in (\ref{randmixt2}) is infinite
dimensional, to create a Gibbs sampler with an a.s. finite number of components,
having as its stationary distribution the latter posterior,
we need to augment (\ref{posterior1}) with the following $T+n$ pairs of auxiliary variables
${\cal V}_T=\{(d_i,N_i):i=-T+1,\ldots,n\}$. $T$ pairs corresponding to the past 
unobserved-observations in $x_{-T+1:0}$, and $n$ pairs corresponding to the true observations 
in $x_{1:n}$. Suppose that the pair of variables $(d_i,N_i)$ is associated with 
the observation $x_i$ (observed or unknown). 
Then, the variable $d_i$ is the infinite mixture allocation variable 
for $x_i$, with $\Pi(d_i\,|\,\l)=\l(1-\l)^{d_i-1}$. Moreover, if jointly $d_i$ 
and $N_i$ are distributed as 
\begin{equation}
\label{dNjoint}
\Pi(d_i,N_i\,|\,\l)\propto (1-\l)^{N_i-1}{\cal I}(d_i\le N_i),
\end{equation}
where ${\cal I}(d_i\le N_i)$ equals to $1$ whenever $d_i\le N_i$ and 0 elsewhere,
then and only then, $d_i$ conditionally on $N_i$, will attain the discrete uniform distribution over the 
random set $\{1,\ldots,N_i\}$, thereby, making the associated (${\cal V}_T$-augmented) Gibbs sampler 
having order $N^*=\max_{-T+1\le i\le n}N_i<\infty$, a.s. 
Finally, when $d_i\le N_i$, because 
$\Pi(x_i\,|\,d_i,\tau_{1:\infty})={\cal N}(x_i\,|\,0,\tau_{d_i}^{-1})$, we have
$$
\Pi(x_i,d_i,N_i\,|\,\l,\tau_{1:\infty})=\l^2(1-\l)^{N_i-1}
{\cal N}(x_i\,|\,0,\tau_{d_i}^{-1}).
$$
In fact, conditioning further the latter density with the variables
$\vt$ and $x_{i-d:i-1}$, yields 
\begin{align}
\label{xidiNi}
& \Pi(x_i,d_i,N_i\,|\,\l,\tau_{1:\infty},\vt,x_{i-d:i-1})=\l^2(1-\l)^{N_i-1}\\
& \quad\quad\quad\times\,{\cal N}(x_i\,|\,g(\vt,x_{i-d:i-1}),\tau_{d_i}^{-1})
\,\,\,{\rm when}\,\,\,d_i\le N_i.\nonumber
\end{align}

Because the global stable manifold possibly extends to infinity, to make the 
Gibbs sampler more efficient, we truncate the coordinates of the 
past unobserved-observation variables $x_{-T-d+1:-T}$ and
$x_{-T+1:0}$ to a finite set $\cal X$ centered at the origin. In that fashion:
\begin{enumerate}
\item To the $T$-shifted to the past initial condition variable $x_{-T-d+1:-T}$ 
      we will assign the uniform prior over ${\cal X}^d$, i.e. 
      $x_{-T-d+1:-T}\sim{\cal U}({\cal X}^d)$, equivalently
      \begin{equation}
      \label{ICprior}
      \Pi(x_{-T-d+1:-T})\propto\prod_{1\le i\le d}{\cal I}(x_{-T-d+i}\in{\cal X}).
      \end{equation}

\item To the variables intermediate between $X_{-T-d+1:-T}$ and and the 
      observed time-series $x_{1:n}$, we apply the restriction event
      ${\cal R}_T=\{x_{-T+1:0}\in{\cal X}^T\}$. 
      Then, it is clear that
      \small{  
      \begin{equation}
      \label{Rintermediate}
      \Pi(x_{-T+1:0}\,|\,{\cal R}_T) \propto \Pi(x_{-T+1:0})
      \prod_{1\le i\le T}{\cal I}(x_{-T+i}\in{\cal X}). 
      \end{equation}      
      }      
\end{enumerate}

We have the following proposition:\\
\vspace{0.025in}\noindent
{\bf Proposition $1$.}
{\it The ${\cal R}_T$-restricted and ${\cal V}_T$-augmented posterior incorporating 
     the $T$-shifted to the past initial condition variable $x_{-T-d+1:-T}$,
     under the assumption that a-priori and independently, $x_{-T-d+1:-T}\sim{\cal U}({\cal X}^d)$,
     is proportional to: 
\begin{equation}
\label{propo1}
\Pi(\l)\,\Pi(\tau_{1:\infty})\,\Pi(\vt)\prod_{-T+1\le j\le d} {\cal I}(x_{j-d}\in{\cal X})
\prod_{\substack{-T+1\le j\le n\\ d_j\le N_j}}\l^2(1-\l)^{N_j-1}
{\cal N}(x_j\,|\,g(\vt,x_{j-d:j-1}),\tau_{d_j}^{-1}).
\end{equation}      

The proof is given in the Appendix A.
}

\vspace{0.025in}
\noindent
{\bf Completing the model:}~
Finally, we have to assign independent priors over the variables
$\tau_{1:\infty}$ and $\vt$. Following standard Bayesian modeling, to the precisions 
$\tau_i$ we apply the conjugate gamma priors $\tau_i\iid{\cal G}(b_1,b_2)$, $i\ge 1$, 
with fixed shape and rate hyperparameters, $b_1$ and $b_2$,
respectively. Although there is an infinity of them, only at most $N^*<\infty$ of them will
be used in the posterior computations. 
We model the deterministic part $g$ of $\cal T$, with the complete
quadratic polynomial in the two variables, given by
\begin{align}
g(\vt,x_{i-2},x_{i-1})=\vt_0+\vt_1x_{i-1}+\vt_2x_{i-2}+\vt_3x_{i-1}x_{i-2}+\vt_4x_{i-1}^2+\vt_5x_{i-2}^2.\label{model}
\end{align}
Over $\vt$, we place the prior $\Pi(\vt)\propto 1$.
Such a prior, although improper, is acceptable due to the fact that it leads to a proper
posterior.   

\vspace{0.025in}
At each iteration of the Gibbs sampler, we  we will sample from the full
conditional densities of the variables:
\begin{align}
&  \tau_{1:N^*},\,d_{-T+1:n},\,N_{-T+1:n}\nonumber\\
&  X_{-T-d-1:-T},\,X_{-T+1:0},\,\vt,\,\l,\,e_{n+1}.\nonumber
\end{align}

It can be verified via (\ref{propo1}), that the full 
conditional densities of the individual variables in the $T$-shifted to the past 
initial condition $x_{-T-d+i:-T}$, and the intermediate part $x_{-T+1:0}$, are given by:
\small
\begin{align}
\label{FC1}
&   f(x_{-T-d+j}\,|\cdots)\propto{\cal I}\left(x_{-T-d+j}\in{\cal X}\right)
    \exp\left\{-{1\over 2}\sum_{1\le l\le j}\tau_{d_{-T+l}}
    \left[x_{-T+l}-g(\vt, x_{-T+l-d:-T+l-1})\right]^2\right\},\,1\le j\le d,\\
\label{FC2}
&   f(x_{-T+j}\,|\cdots)\propto{\cal I}\left(x_{-T+j}\in{\cal X}\right)
    \exp\left\{-{1\over 2}\sum_{0\le l\le d}\tau_{d_{-T+j+l}}
    \left[x_{-T+j+l}-g(\vt, x_{-T+j+l-d:-T+j+l-1})\right]^2\right\},\,1\le j\le T,
\end{align}
\normalsize
respectively, where $f(x\,|\cdots)$ denotes the density of $x$ conditional on 
the rest of the variables. 

We remark that, although both previous collections of densities are modeling past out-of-sample observations, the full conditionals associated with the $T$-shifted to the past starting point in (\ref{FC1}), exhibit the unique characteristic that they jointly depend
purely on future observations, in the sense that: 
\begin{align}
& {\rm starting~seq.:}~f(x_{-T-d+1:-T}\,|\,x_{-T+1:-T+d},\cdots)\nonumber\\
& {\rm intermediate~seq.:}~f(x_{-T+1:0}\,|\,x_{-T-d-1:-T},x_{1:d},\cdots).\nonumber
\end{align}
All other variables, have full conditionals that depend equally on both future and past observations. This apparent lack of information, makes the support of the associated full conditional diffusive along the direction of the local stable manifold contained on a  neighborhood of the true (but unknown) initial point $x_{-d+1:0}$ of the observed 
time-series $x_{1:n}$.

%===========================================================================================	

\subsection{The stable manifold stochastic approximation}

The proposed method is based on the observation that the posterior marginal distribution of the initial condition vector lies along the \textit{stable direction}, i.e. the direction of the stable manifold \cite{hao1998applied}. We use the stable direction, for example in order to find regions of HTs, by iterating a random normalized vector under the inverse Jacobian matrix at a given point\cite{christiansen2013future, lai1993often}. In practice, due to the influence of numerical errors, we use the evaluation of the inverse Jacobian at each orbit point. More details regarding the calculation of stable and unstable directions, can be found in \cite{lai1993often}. 

In all our numerical illustrations, we will attempt a completely noninformative prior 
specification, namely we will set:
\begin{equation}
\label{specif1}
\l\sim{\cal B}e(\,\cdot\,|\,0.5,0.5),\,\,
\tau_{1:\infty}\sim\prod_{i=1}^\infty{\cal G}(\,\cdot\,|\,10^{-3},10^{-3}).
\end{equation}
We remark that $\l$ a-priori, follows the arcsine density, coinciding 
with the Jeffrey's prior. The prior over the independent inverse variances $\tau_i$
is a vague gamma prior, being very close to a scale invariant prior.

\vspace{0.03in}
\noindent
{\bf The multiple time-series approach:~}
We will now set up a basic illustrative numerical example, concerning the support of
the T-shifted to the past initial points for the H\'enon map. We will consider 
the sequence of $r=250$ initial points ${\cal C}_g^r=\{(y_j,y_{j+1}):1\le j\le r\}$, 
generated by means of the H\'enon map
$$
y_j = g(y_{j-2},y_{j-1}) = 1.38 - y_{j-1}^2 + 0.27y_{j-2},
$$
with starting point $y_{-1:0}=(-0.61,1.37)$. Next, for each initial point
$x_{-1:0}^{(j)}\in{\cal C}_g^r$, $j=1,\ldots,r$, we generate an associated $f$-noisy time-series $x_{1:n}^{(j)}$ of length $n=500$, via 
\begin{align}
&  x_i^{(j)} = g(x_{i-2}^{(j)},x_{i-1}^{(j)})+e_i^{(j)},\nonumber\\
\label{random_Henon_1}
&  e_i^{(j)}\iid f={9\over 10}{\cal N}(0, 10^{-7})+{1\over 10}{\cal N}(0, 10^{-3}).
\end{align}
Now, conditional on the $r$ ``observed'' $f$-noisy time-series $x_{1:n}^{(j)}$, we sample
from the posterior marginal distribution of the T-shifted to the past 
initial point variable $X_{-T-d+1:-T}$. 
We restrict the intermediate variables $X_{-1}$ and $X_0$ in the posterior (\ref{propo1})
over the interval ${\cal X}=(-2,2)$, and we set a-priori 
$(X_{-3},X_{-2})\sim{\cal U}({\cal X}^2)$.
Finally, we consider the union of the supports 
of the variable $(X_{-T-d+1:-T}\,|\,x_{1:n}^{(j)})$ for $j=1,\ldots,r$, as an 
$(r,n)$-approximation to the perturbed global manifold of $g$; so that in the
multiple time-series approach the SM-GSBR approximation is given by:
$$
{\hat W}^s(T;r,n)=\bigcup_{1\le j\le r}{\rm supp}\left(X_{-T-d+1:-T}\,|\,x_{1:n}^{(j)}\right). 
$$

In Fig. 4(a) we display the set of initial conditions ${\cal C}_g^r$ in black small
circles. Together, we superimpose the set of points in ${\hat W}^s(T=0;r,n)$ in red. This is
the union of the supports of the initial points $(X_{-1}^{(j)},X_0^{(j)})$
corresponding to the time-series $x_{1:n}^{(j)}$, for $j=1,\ldots,r$. Together
we superimpose the set of points ${\hat W}^s(T=2;r,n)$ in gray.
We can see how the increase of the backward prediction horizon to $T=2$,
drives the supports of the posterior distributions along the stable direction. 
In Fig. 4(b)-(c) we provide enlargements of certain regions of the state space. 

\begin{figure}[hbt!]
	\centering
	\includegraphics[width =0.85\textwidth]{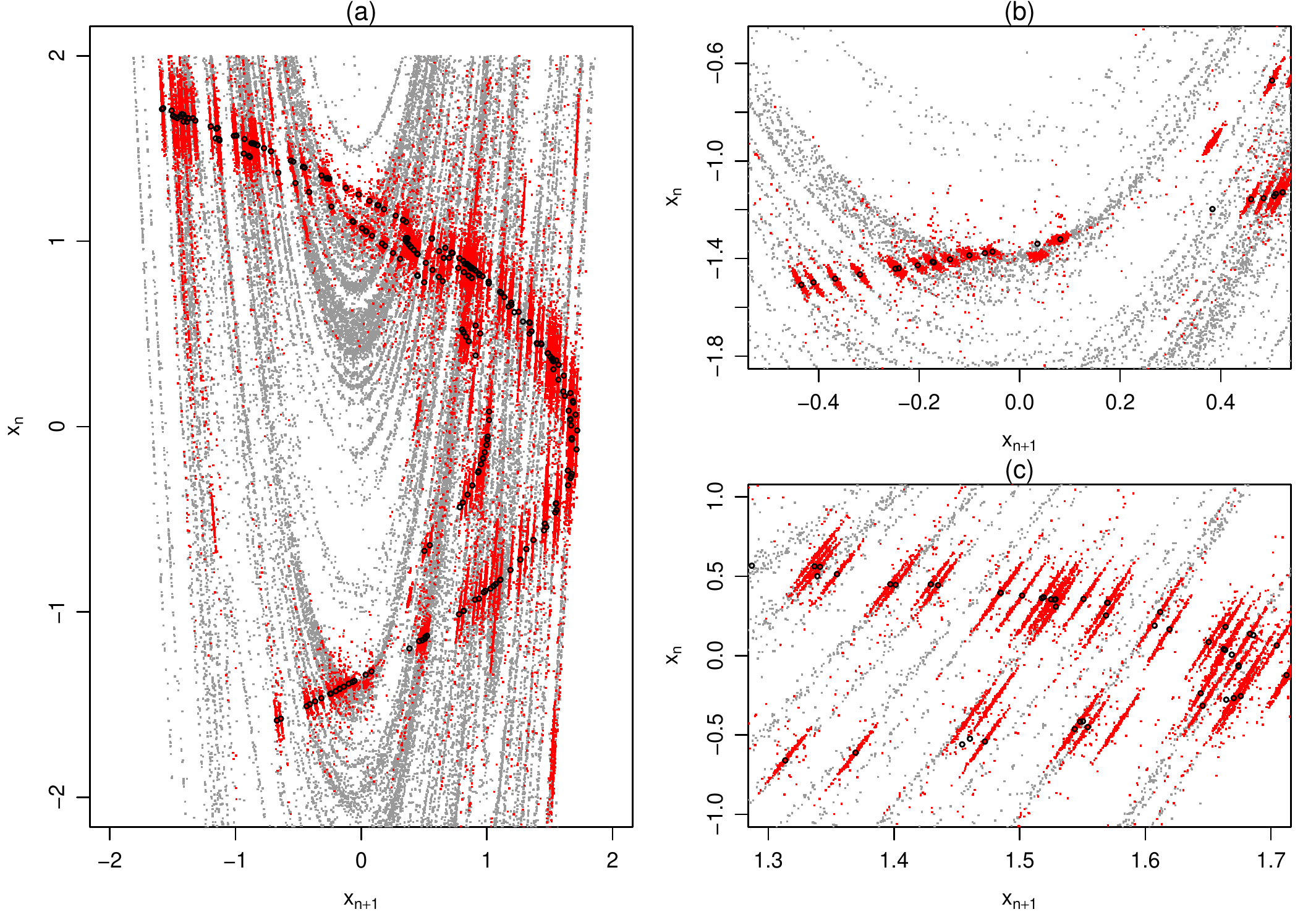}
	\caption{In Fig. (a) the initial conditions in the set ${\cal C}_g^r$ are displayed in black. 
	         The union of predictive samples in ${\hat W}^s(T=0;r,n)$ and
	         ${\hat W}^s(T=2;r,n)$, are depicted in red and gray respectively. 
           Enlargements of the regions positioned at $[-0.50,0.55]\times[-1.8,-0.5]$ and
           $[1.30,1.75]\times[-1, 1]$, are provided in Figs (b) and (c), respectively. 
           \label{x0y0t0t21}}
\end{figure}

We remark that the quality of the initial points estimations $\hat{x}_{-1:0}^{(j)}$,
associated with the time series $x_{1:n}^{(j)}$, for $j=1,\ldots,m$,
depend not only on the perturbation realizations $e_{1:n}^{(j)}$, but also on the true
position of the initial points in the state space. 

\begin{figure}[hbt!]
	\centering
\includegraphics[width=0.6\textwidth]{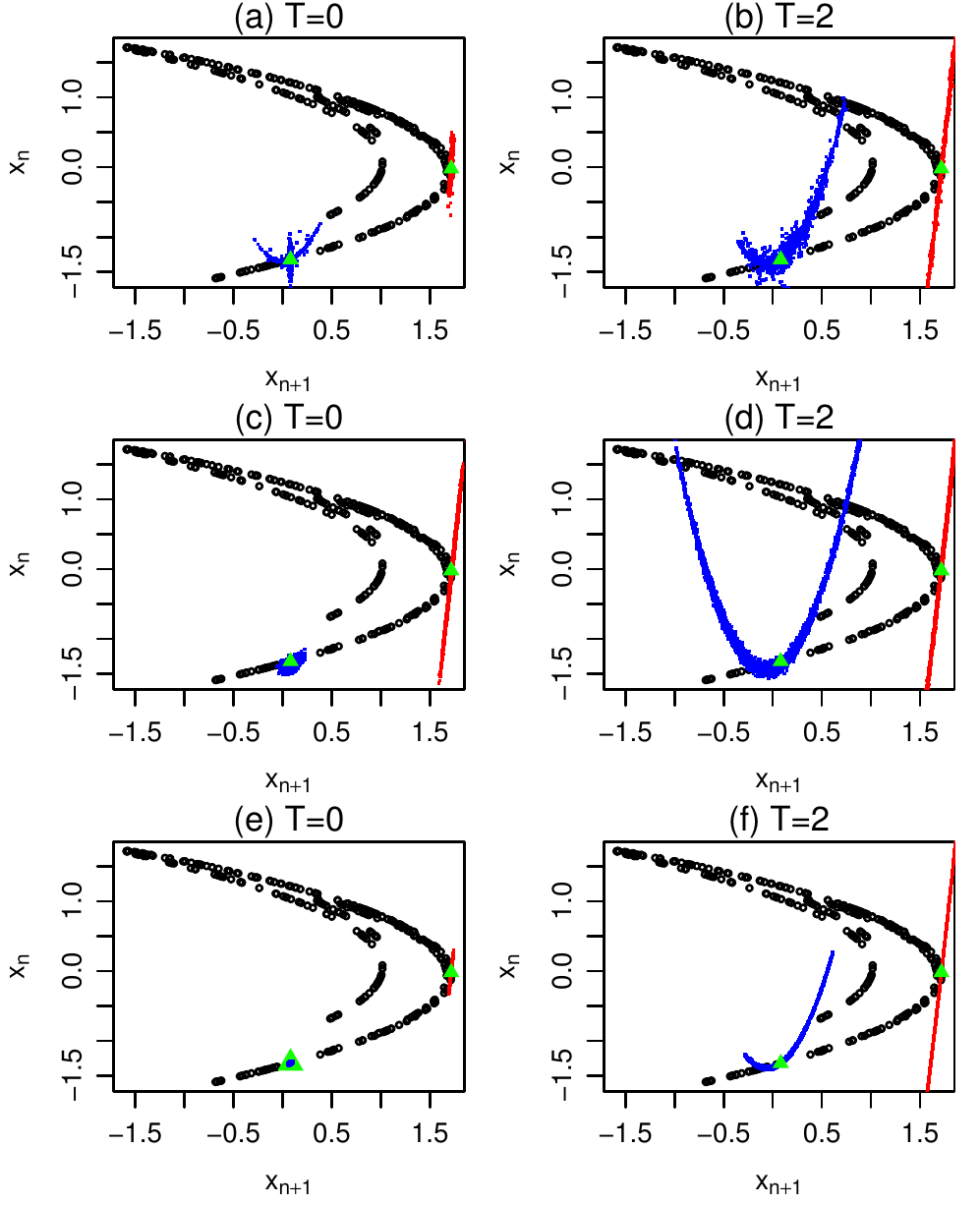}
\caption{The embedded deterministic orbit in ${\cal C}_g^{250}$ (in black), 
         is superimposed with the predictive samples of the initial points for
         $T=0$ and $T=2$, near the points $(0.08,-1.32)$ (in blue) and $(1.71,-0.02)$ 
         (in red). The figures in the first row correspond to the non-Gaussian noise in 
         (\ref{random_Henon_1}). Second row figures, correspond to a Gaussian noise
         perturbed time-series of variance equal to that of the non-Gaussian noise.
         The last row corresponds to the zero noise limit. \label{2x0y0}}
\end{figure}

In Fig. \ref{2x0y0} we depict the sequence of points
in ${\cal C}_g^s$ by small black circles. Together, we superimpose the predictive samples of the posterior marginal initial conditions of two special initial points that are in the neighborhood of primary HTs, for $T=0$ and $T=2$, in Fig. \ref{2x0y0}(a),(c),(e) 
and Fig. \ref{2x0y0}(b),(d),(f), respectively. 
Near the initial point $x_{-1:0}^{(87)}=(0.08,-1.32)$ (in blue) the stable manifold exhibits locally relatively high curvature, while near the initial point 
$x_{-1:0}^{(150)}=(1.71,-0.02)$ (in red) the curvature of the local stable manifold 
is close to zero. More specifically, a-posteriori the initial point variables
have their support very close to the local stable direction, so that 
the structure of the support depends on the curvature of the local stable manifold. 
In Fig. \ref{2x0y0}(a)-(b) the posterior computations are based on the synthetic
time-series produced by the non-Gaussian noise process in (\ref{random_Henon_1}).
In Fig. \ref{2x0y0}(c)-(d) the posterior computations are based on time series 
produced with ${\cal C}_g^r$ with Gaussian noise, of variance equal to the variance 
of the non-Gaussian noise. Finally, in Fig. \ref{2x0y0}(e)-(f) we have employed the corresponding deterministic time-series. Clearly, the most fuzzy results qualify 
the non Gaussian perturbations.

\vspace{0.025in}
\noindent{\bf The sliding window time-series approach:~}
In practice, most of the times we have only one available data set $x_{1:n}$, generated by some unknown random dynamical system. In order to reveal the macroscopic structure of the stable 
manifold, we apply the T-backward prediction algorithm a multiple of times, each time based on a sliding window time-series subset of the observed time-series. 

Therefore, after restricting the intermediate variable 
$X_{-T+1:0}$ in the posterior (\ref{propo1}) over ${\cal X}^T$ and assigning to the variable $X_{-T-d+1:-T}$ the  uniform prior over ${\cal X}^d$, 
conditionally on the $k$ sliding window time-series $x_{j:n-k+j}$ of length
$n-k+1$, we sample from the posterior marginal distribution of the T-shifted 
to the past initial point variable $X_{-T-d+1:-T}$. So, based on a single observed time-series,
we define the SM-GSBR approximation to the perturbed stable manifold of $g$ as the union of all the associated supports of the variables $(X_{-T-d+1:-T}\,|\,x_{j:n-k+j})$ for all $j=1,\ldots,k$, namely
\begin{equation}
\label{singleBGSBR}
{\hat W}_n^s(T;k)=\bigcup_{1\le j\le k}{\rm supp}
\left(X_{-T-d+1:-T}\,|\,x_{j:n-k+j}\right). 
\end{equation}

Regarding the choice of the prediction horizon $T$, we have observed that small values of 
$T=2,3,4$, are adequate. 
%In fact, lower values of $T$ should preferred, whenever computational resources are limited.

%In Fig. \ref{fig:x0sprop} we have set the initial point to $\left(x_{-1},x_{0}\right) = (-2,0)$ and we have generated a sample $\left(x_1,\ldots,x_n\right)$ of length $n=500$, with the standard parameters of the H\'{e}non map, using additive Gaussian noise with variance $\s^2 = 10^{-7}$. We have applied the B--GSBR model, for different reversed time horizons $T=0,1,2,3$. 
%Each time, the initial condition was the same, so when $T=i, i=0,1,2,3$ we have used the $\left(x_{T+1},x_{T+2},\ldots,x_{n}\right)$ subset of the original data set. 

%\begin{figure}
%	\centering
%	\includegraphics[width =0.5\textwidth]{Figures/x0spropeps}
%	\caption{Embedded data from noisy H\'{e}non map (black), superimposed with the posterior marginals of the initial condition vector (red) and past unobserved observations (gray). True values of initial condition and true past observations are indicated with square and circles respectively .\label{fig:x0sprop}}
%\end{figure}

\section{Simulation results}\label{s4}

In this section, we will provide numerical illustrations of the SM-GSBR algorithm, in approximating the global stable manifold of the H\'{e}non and the Dual-H\'{e}non maps, using single synthetic time series. Moreover, we will apply the SM-GSBR model on the orbits of a H\'{e}non type noninvertible polynomial map.

In all our numerical experiments, we use the noninformative prior specifications
(\ref{specif1}) and the model polynomial map in (\ref{model}). We will set the 
restriction-prior interval $\cal X$ in posterior (\ref{propo1}) on a case-by-case 
basis. The SM-GSBR Gibbs 
sampler executions ran for $20\times 10^4$ iterations, leaving the first 
$5\times 10^4$ samples as a burn-in period. Thinning has been performed 
in every 150 iterations as a means for reducing the correlation between sampled values.

\subsection{The H\'{e}non map}

\vspace{0.025in}
\noindent{\bf The zero noise limit case:~}
In this case, the observed time-series $x_{1:n}$ is deterministic
of size $n=2000$, coming from the classical H\'{e}non map
\begin{equation}
\label{classicalHenon}
x_i=1-1.4\,x_{i-1}^2+0.3\,x_{i-2}, 
\end{equation}
with initial point at $(-1,0.5)$. In fact, we consider the 
deterministically generated time-series $x_{1:n}$, as being additively corrupted by a 
close to zero intensity, unknown dynamical noise process $f^*$. For example, we could assume the existence of such a zero noise limit component $e_i\sim f^*$ as coming from the errors produced at each iteration of the map, caused by the finite precision of the floating point computer arithmetic.

In this case, we have fixed the time reversed prediction horizon to $T=3$ with a
$k=500$ sliding window time-series. We restrict the intermediate variables 
$X_{-2},X_{-1}$ and $X_0$ over the interval ${\cal X}=(-3,3)$, and we set a-priori 
$(X_{-4},X_{-3})\sim{\cal U}({\cal X}^2)$. The resulting SM-GSBR approximation 
of the stable manifold in (\ref{singleBGSBR}) is given by
$$
{\hat W}_{2000}^s=\bigcup_{1\le j\le 500}{\rm supp}(X_{-4},X_{-3}\,|\,x_{j:j+1500}). 
$$

\begin{figure}[hbt!]
	\centering
	\includegraphics[width =0.75\textwidth]{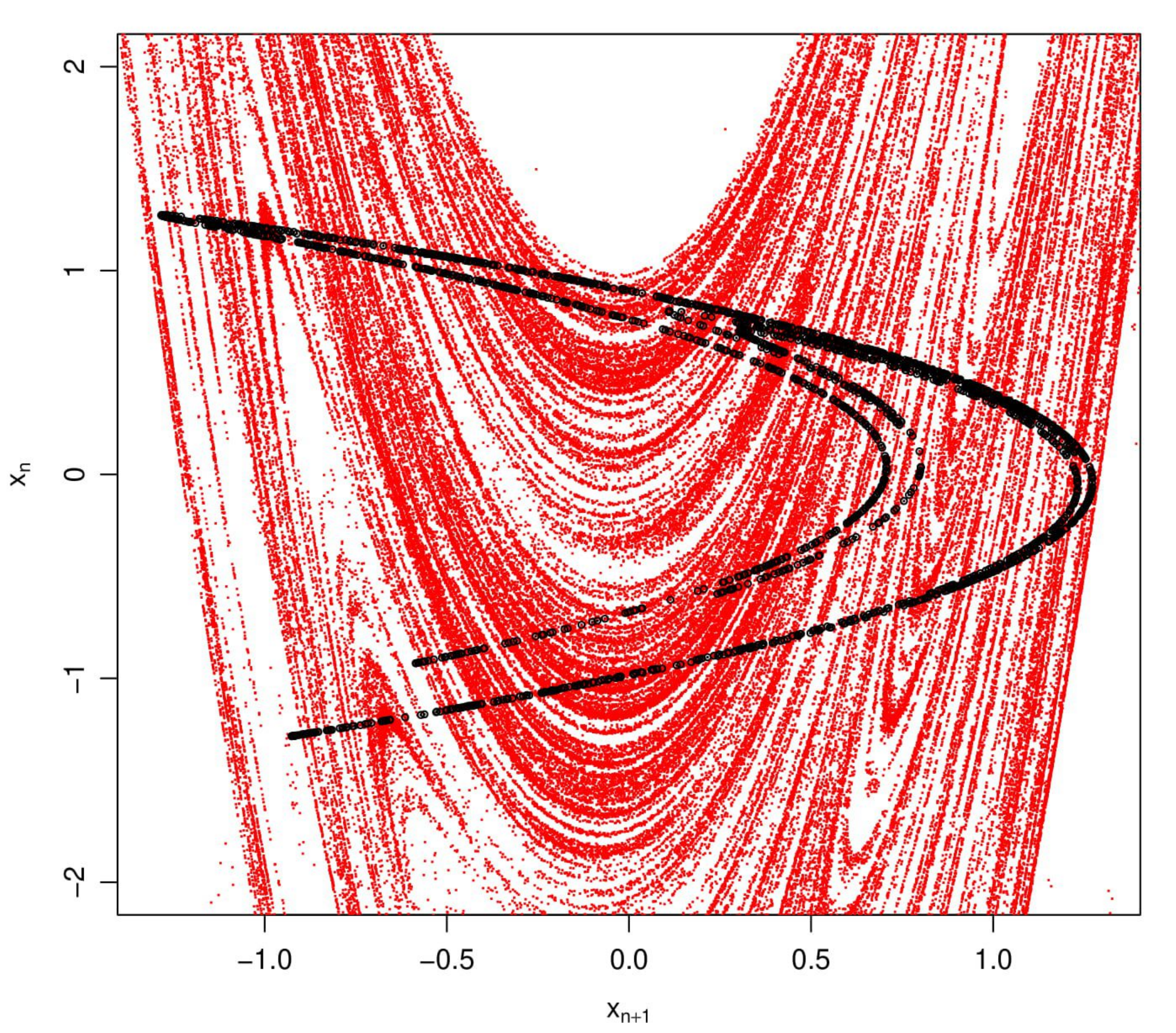}
	\caption{The $\R^2$-embedded deterministic H\'{e}non time-series $x_{1:n}$ is rendered
	         in black. The SM-GSBR predictive sample in ${\hat W}_{2000}^s$, approximating the 
	         global stable manifold, is presented in red. \label{fig:henon1}}
\end{figure}

In Fig. \ref{fig:henon1} we present the $\R^2$-embedded data points lying close to the H\'{e}non attractor in black, superimposed with the sampled values in the set 
${\hat W}_{2000}^s$ in red. 

It is evident that the union of the supports of the joint posterior marginals forms a stochastic approximation of the associated global stable manifold of the deterministic map, as shown in Fig. \ref{fig:henonsm}.

\begin{figure}[hbt!]
	\centering
	\includegraphics[width =0.75\textwidth]{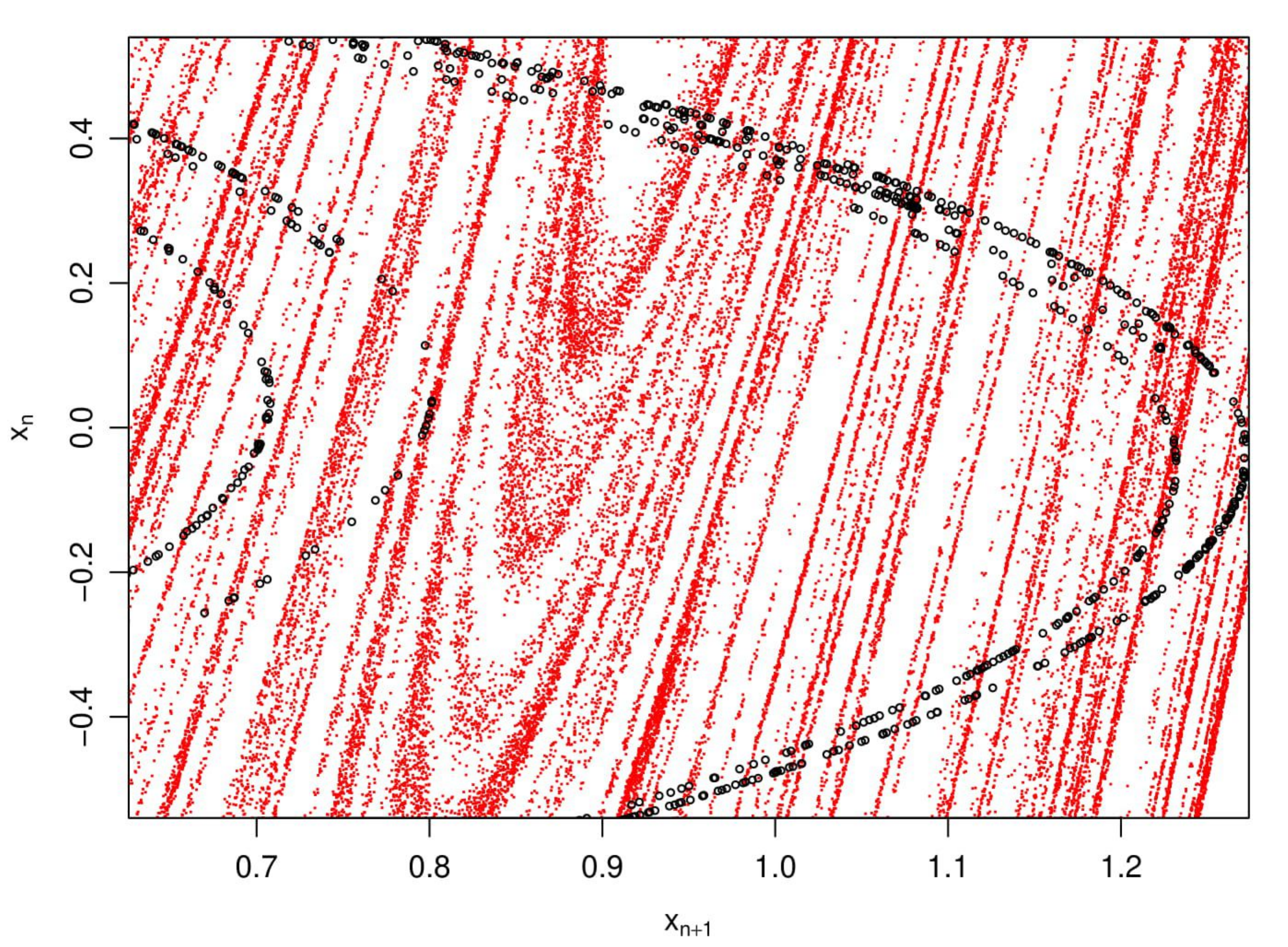}
	\caption{Enlargement of the rectangle $\left[0.65,1.25\right] \times \left[-0.5,0.5\right]$ of Fig. \ref{fig:henon1}. \label{fig:henon2}}
\end{figure}

In order to further investigate the quality of the stochastic approximation, we show in Fig. \ref{fig:henon2} data points and sampled values from the joint posterior marginals over the rectangle $\left[0.65,1.25\right] \times \left[-0.5,0.5\right]$.

\vspace{0.025in}
\noindent{\bf Non-Gaussian noise:~}
Here we generate a time series $x_{1:n}$ of length $n=1000$, using the H\'{e}non map 
in (\ref{classicalHenon}), with the same initial point, yet with an additive stochastic component which is given by
$$
e_i\iid f={3\over 4}{\cal N}(0, \s_1^2)+{1\over 4}{\cal N}(0, \s_2^2),
$$
with $\s_1^2=10^{-6}$ and $\s_2^2=10^{-3}$. 

We set $T=3$, the number of sliding time-series to $k=250$, and the restriction-prior interval to ${\cal X}=(-6,2)$. We denote the SM-GSBR approximation of the global stable manifold by
$
{\hat W}_{1000}^s=
\cup_{1\le j\le 250}\,{\rm supp}\left(X_{-4},X_{-3}\,|\,x_{j:j+750}\right).
$

We note that due to the presence of corrupting noise, the approximation of the global stable manifold is subjected to a blurring effect. However, with the proposed SM-GSBR method, we are still able to gain insight about the qualitative characteristics of the dynamical behavior of the underlying system responsible for the observed data set.

In such cases, instead of presenting the approximation to the stable manifold as the union of supports of posterior marginals, it will be more instructive to present the approximation, 
with respect to higher posterior density regions (HPDRs) supported over the truncating set 
$\cal X$. More specifically, we apply a high analysis grid over $\cal X$, and we color-code 
regions according to the frequency of the predictive samples in ${\hat W}_{1000}^s$.

\begin{figure}[hbt!]
	\centering
	\includegraphics[width =0.75\textwidth]{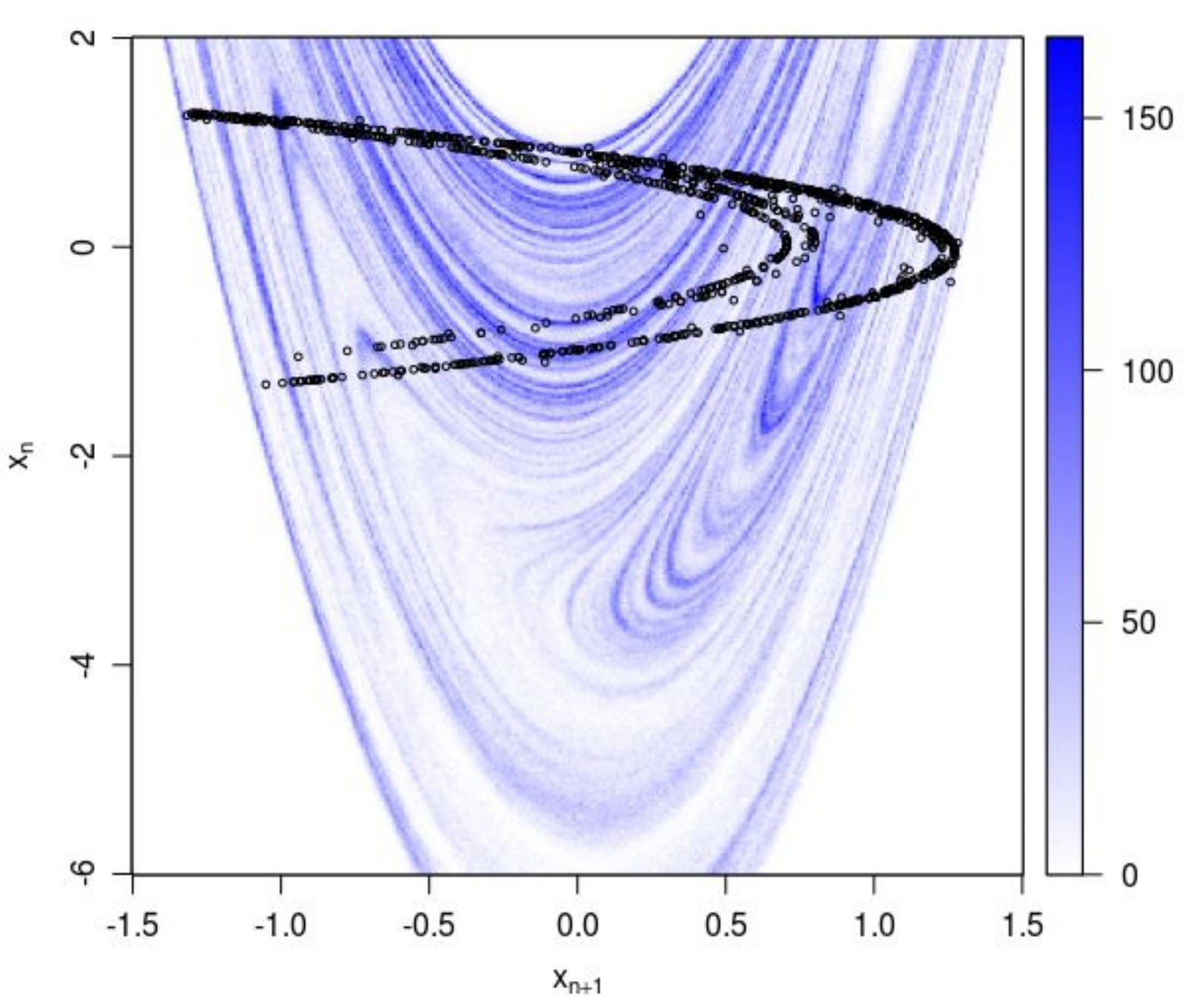}
	\caption{The $\R^2$-embedded noisy time-series is depicted in black. 
	         The HPDR, based on the SM-GSBR predictive sample ${\hat W}_{1000}^s$, 
	         is superimposed using a high analysis grid. 
	         \label{fig:henon3}}
\end{figure}

In Fig. \ref{fig:henon3} we present the $\R^2$-embedded noisy time-series $x_{1:n}$ of length 
$n=1000$ in black. Together, we superimpose the HPDR set based on the SM-GSBR approximation. 

\subsection{The Dual-H\'{e}non map}

\vspace{0.025in}
\noindent{\bf The zero noise limit case:~}
In this subsection we will consider the Dual-H\'{e}non map
\begin{equation}
\label{GHM1}
x_i=2\,x_{i-1}-0.1\,x_{i-1}^3+0.3\,x_{i-2},
\end{equation}
which is a special case of the Generalized H\'{e}non map in (\ref{dualhenon}).
These is the case in which, two symmetric with respect to the origin, isolated 
H\'{e}non-like attractors ${\cal A}^+$ and ${\cal A}^-$ coexist (see Fig. 5.3),
located in the first and third quadrant, respectively.

We generate two deterministic orbits $x_{1:n}^+$ and $x_{1:n}^-$ of length
$n=1000$, by iterating the initial points $p^+=(1,0.5)$ and $p^-=(0.5,1)$, 
lying on the basins of attraction of ${\cal A}^+$ and ${\cal A}^-$, respectively. 

We use the two orbits as observed data sets, and to each orbit, we apply the 
SM-GSBR sampler with reversed time horizon $T=2$, with $k=500$ sliding time-series.
We set the restriction-prior interval to ${\cal X}=(-5,5)$, so that a-priori
$(x_{-3},x_{-2})\sim{\cal U}({\cal X}^2)$, and the variable $(X_{-1},X_0)$ 
is truncated over the square ${\cal X}^2$. The two corresponding SM-GSBR approximations
are given by
$$
{\hat W}_{1000}^{s\,\pm}=
\bigcup_{1\le j\le 500}{\rm supp}(X_{-3},X_{-2}\,|\,x_{j:j+500}^\pm).
$$

\begin{figure}[hbt!]
	\centering
	\includegraphics[width =0.75\textwidth]{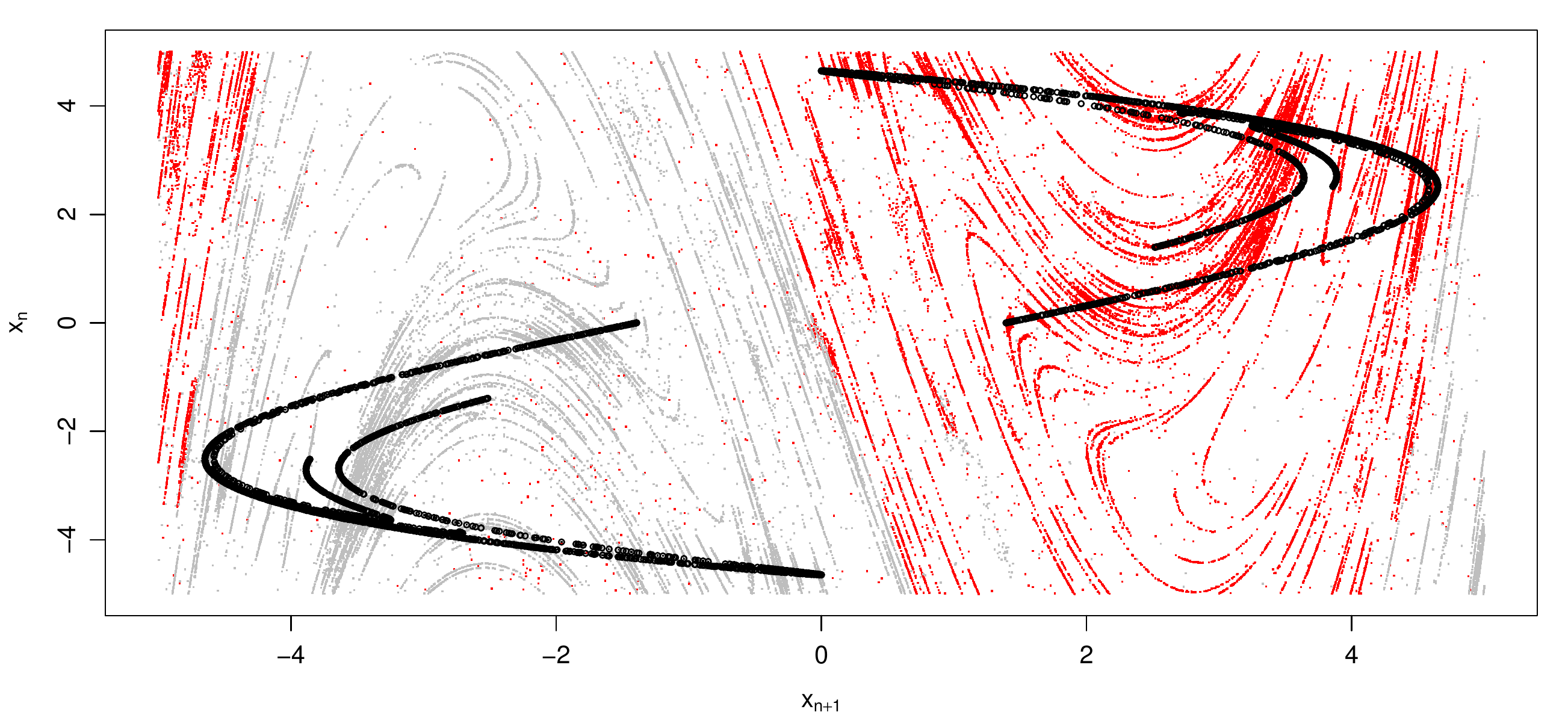}
	\caption{The $\R^2$-embedded data sets $x_{1:n}^+$ and $x_{1:n}^-$ of the deterministic
	         Dual-H\'{e}non map, are depicted in black. 
	         Together are superimposed the two SM-GSBR approximations ${\hat W}_{1000}^{s\,+}$ 
	         and ${\hat W}_{1000}^{s\,-}$ of the two stable manifolds, in red and gray, respectively.
	         \label{fig:dhenon1}}
\end{figure}

In Fig.\ref{fig:dhenon1} we present the $\R^2$-embedded data sets $x_{1:n}^+$ 
and $x_{1:n}^-$ (in black), lying close to the coexisting strange attractors
${\cal A}^+$ and ${\cal A}^-$, respectively. Together we superimpose the  
SM-GSBR approximations ${\hat W}_{1000}^{s\,+}$ and ${\hat W}_{1000}^{s\,-}$ of the two 
stable manifolds emanating from the two coexisting saddle fixed points (see Fig. \ref{fig:dhsm}), in red and gray, respectively.

We remark, that the two bivariate densities $\Pi(x_{-3},x_{-2}\,|\,x_{1,n}^\pm)$,
for computational simplicity have common support $\cal X$. This is the reason 
why a very sparse cloud of points (in red) belonging to ${\hat W}_{1000}^{s\,+}$ can be observed 
inside the basin of attraction of ${\cal A}^-$ and vise-versa.

At this point, we would like to emphasize the fact that a-priori, there was no information 
on the control parameters of the map or the location of the saddle fixed points. Our 
stochastic approximation, given the general functional representation of the deterministic 
part is completely driven by the observed time-series.

\vspace{0.025in}
\noindent{\bf The case of an impulsive noise process:~}
Here we mostly elaborate on the efficiency of the proposed SM-GSBR model.
We generate a short time-series $x_{1:n}^+$ of length $n=500$, via the Dual-H\'{e}non map 
in (\ref{GHM1}), yet influenced by the additive impulsive 
stochastic component 
\begin{equation}
\label{impulsive1}
e_i\iid f={9\over 10}{\cal N}(0, \s_1^2)+{1\over 10}{\cal N}(0, \s_2^2),
\end{equation}
with $\s_1^2=10^{-7}$ and $\s_2^2=10^{-2}$. The initial condition of the noisy
orbit $x_{1:n}^+$ is given by $p^+=(1, 0.5)$.

We set $T=2$, with the number of sliding time-series decreased to $k=150$. 
At the same time we expand the restriction-prior interval to ${\cal X}=(-10,10)$. 
We denote the associated SM-GSBR approximation by
$$
{\hat W}_{500}^{s\,+}=
\bigcup_{1\le j\le 150}\,{\rm supp}\left(X_{-3},X_{-2}\,|\,x_{j:j+350}\right).
$$

\begin{figure}[hbt!]
	\centering
	\includegraphics[width =0.75\textwidth]{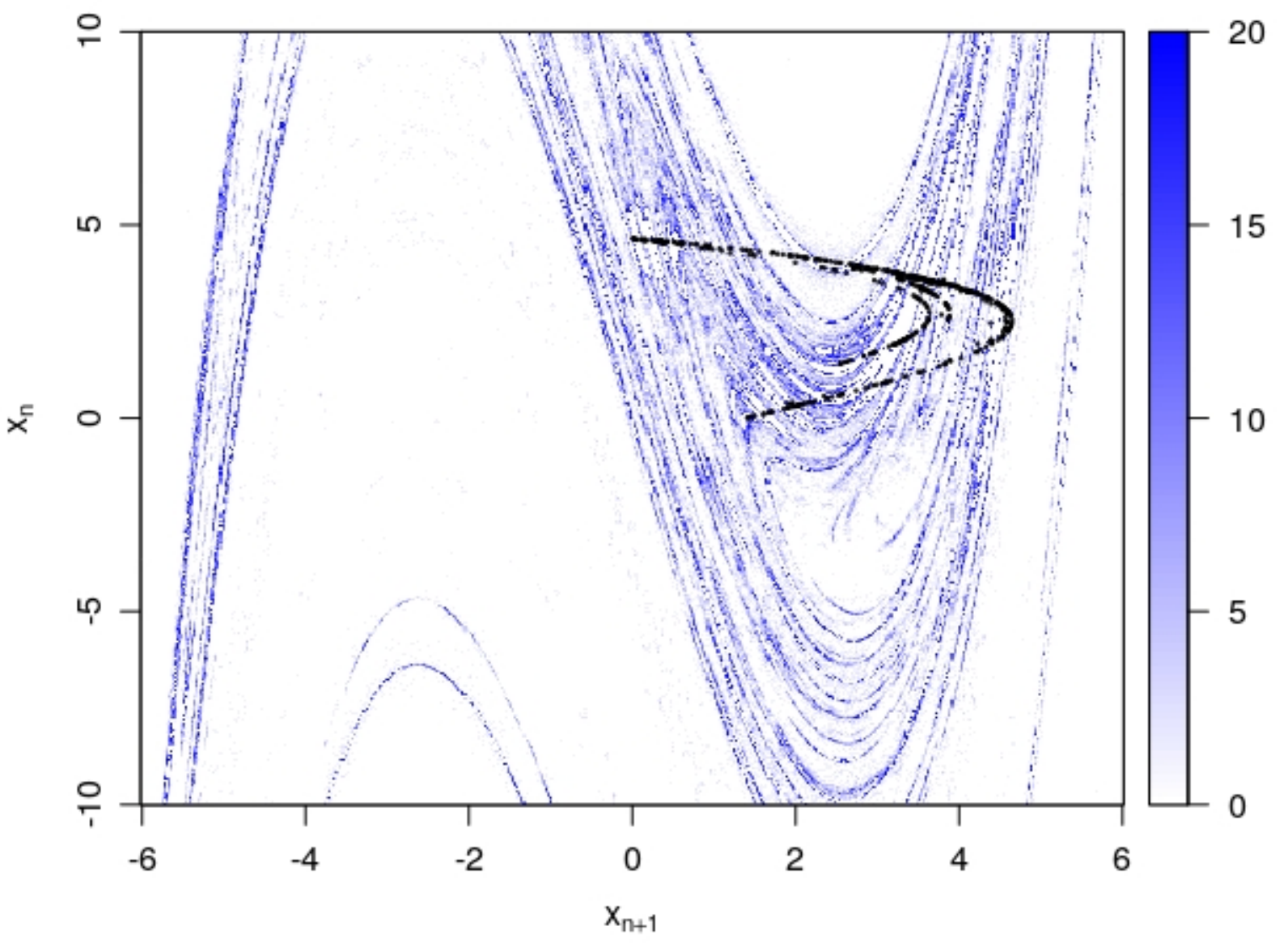}
	\caption{Stochastic approximation of the global stable manifold of the Dual H\'{e}non map. Together we superimpose a HPDR color map of the predictive samples in ${\hat W}_{500}^{s\,+}$. \label{fig:dhsm2}}
\end{figure}

In Fig. \ref{fig:dhsm2} we present the $\R^2$-embedded noisy time-series $x_{1:n}^+$ 
of length $n=500$ in black. Together, we superimpose the HPDI set corresponding to the SM-GSBR approximation ${\hat W}_{500}^{s\,+}$.

We remark that although the noise processes was non-Gaussian and the observed time-series
was short, we were able to produce an adequate approximation to the stable manifold of the 
saddle fixed point located on the first quadrant.

\subsection{A noninvertible 2-d quadratic map}

In this subsection we illustrate the performance of the SM-GSBR model based
on nonlinear time-series data coming from a noninvertible polynomial map.
For the generation of the synthetic time-series, we make use of the 
noninvertible quadratic map, given by:
\begin{equation}
\label{NIM1}
x_i=1.38 - x_{i-1}^2+0.211\, x_{i-2}^2,
\end{equation}
For a detailed analysis of the dynamical behavior of two--dimensional noninvertible quadratic maps we refer to  \cite{elhadj20102}. 

Since there exist orbit points with more than one preimages, we extend the definition of the stable and unstable manifolds \cite{sander2000homoclinic} to the following:
\begin{align}
&   {\cal S}^s(y)=\{x\in\R^2:\exists\,(x_n)_{n\ge 0},\,\,
    {\rm through}\,\,\,x,\,\,{\rm s.t.}\nonumber\\
&   \qquad\qquad\qquad\qquad\qquad\qquad x_n\to y,\,n\to\infty\},\nonumber
\end{align}
and,
\begin{align}
&   {\cal S}^u(y)=\{x\in\R^2:\exists\,(x_{-n})_{n\ge 0},\,\,
    {\rm through}\,\,\,x,\,\,{\rm s.t.}\nonumber\\
&   \qquad\qquad\qquad\qquad\qquad\qquad x_{-n}\to y,\,n\to\infty\},\nonumber
\end{align}
with $y$ being a fixed point for the map $g$.
Whenever the invertibility of $g$ is not guaranteed, usually ${\cal S}^{s}$ and ${\cal S}^{u}$ are being referred to as the stable and unstable invariant {\it sets}, respectively. The global invariant sets are no longer guaranteed to be manifolds and due to the multiple preimages, they can have intersections, being non--smooth or being totally disconnected \cite{frouzakis2003route}. The approximation of the stable manifold in such case is useful, because it gives us information on the complicated basins of attraction that the multiple preimages created \cite{england2004computing}. The simplest case of noninvertible quadratic maps are of the $(Z_0-Z_2)$ type, meaning that there exist points in the state space with only two or zero preimages, defining the two mutually exclusive basins of attraction.

\vspace{0.025in}
\noindent{\bf The zero noise limit case:~}
We generate the deterministic orbit $x_{1:n}$ of length 2000, via the map in 
(\ref{NIM1}), with initial point at $(0.5,1.5)$. We aim to approximate the 
global stable set of the noninvertible map in (\ref{NIM1}).

We set $T=3$ with $k=500$ sliding time-series and the restriction-prior interval 
to ${\cal X}=(-2,2)$. We denote the SM-GSBR approximation of the stable set over $\cal X$, by
$$
\hat{\cal S}_{2000}^s=
\bigcup_{1\le j\le 500}\,{\rm supp}\left(X_{-4},X_{-3}\,|\,x_{j:j+1500}\right).
$$
  
\begin{figure}[hbt!]
	\centering
	\includegraphics[width =0.75\textwidth]{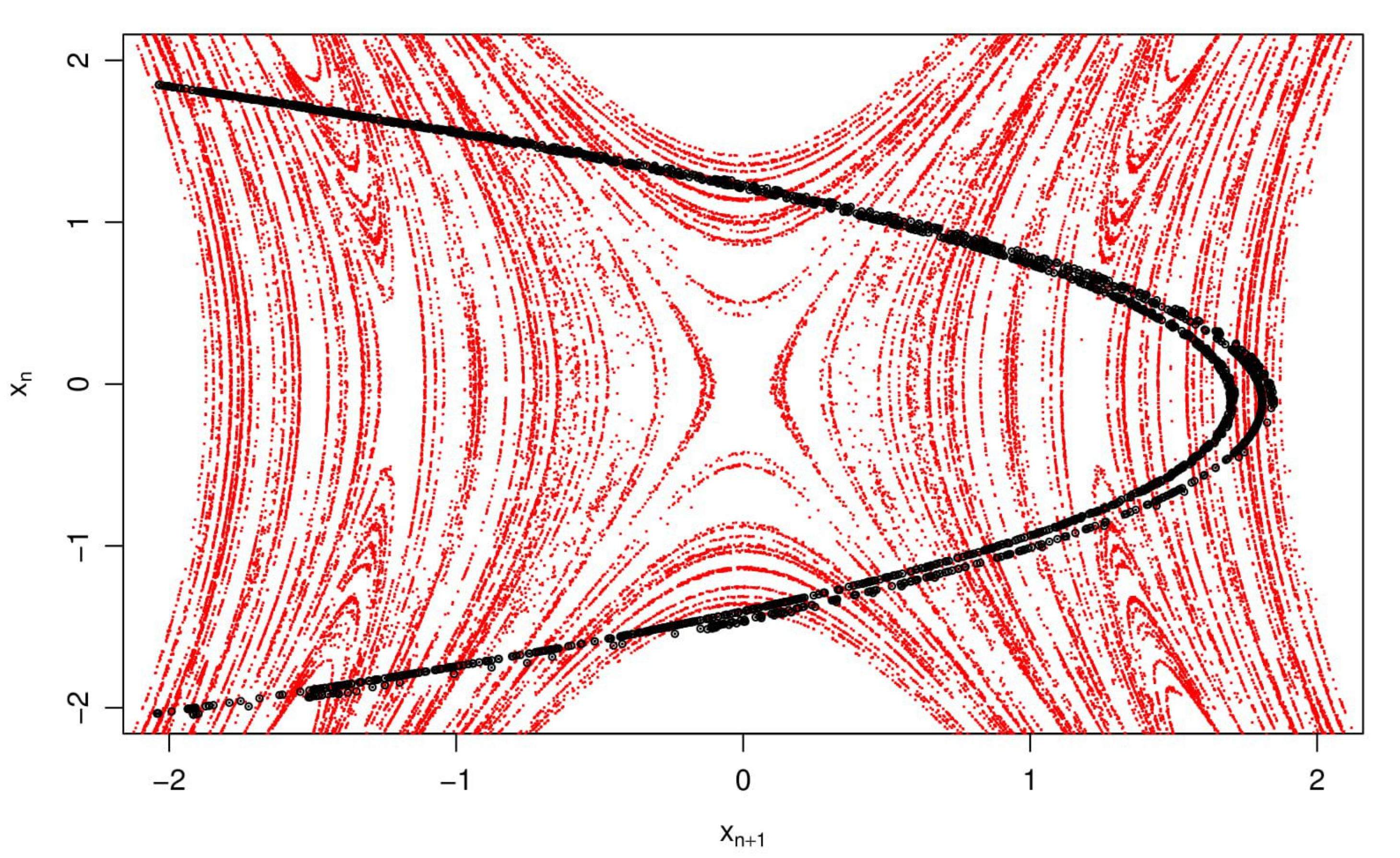}
	\caption{The $\R^2$-embedded deterministic time-seriest $x_{1:n}$ is depicted in black. 
	         Together we have superimposed the union of the predictive sample in 
	         $\hat{\cal S}_{2000}^s$ in red. 
	         \label{figdninv}}
\end{figure}

In Fig. \ref{figdninv} we present the deterministic time-series $x_{1:n}$ in black,
and the union of the predictive sample in ${\hat W}_{2000}^s$ in red.

\vspace{0.025in}
\noindent{\bf The case of an impulsive noise process:~}
We generate a time-series $x_{1:n}$ of length $n=1000$, via the noninvertible map 
in (\ref{NIM1}), influenced by the additive impulsive stochastic component in (\ref{impulsive1}). 

We set $T=2$ with $k=500$ sliding time-series, and we set the restriction-prior interval 
to ${\cal X}=(-4,4)$. We denote the SM-GSBR approximation of the stable set over $\cal X$, by 
$$
\hat{\cal S}_{1000}^s=
\bigcup_{1\le j\le 500}\,{\rm supp}(X_{-3},X_{-2}\,|\,x_{j:j+500}).
$$

\begin{figure}[hbt!]
	\centering
	\includegraphics[width =0.75\textwidth]{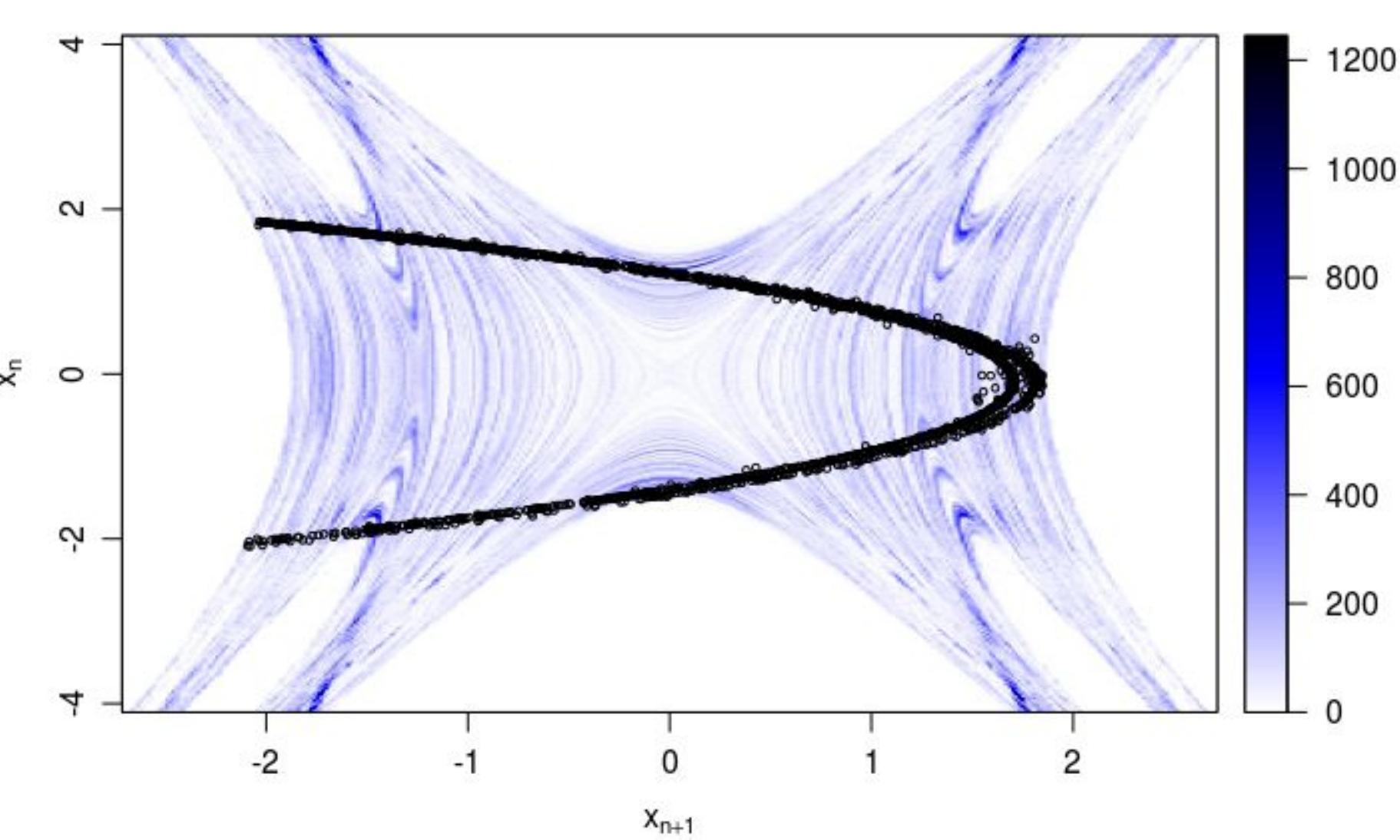}%smrinv,rninvjpg22
	\caption{The $\R^2$-embedded dynamically perturbed time-series $x_{1:n}$ is depicted in black. 
	         Together we superimpose a HPDR color map of the predictive samples in 
	         $\hat{\cal S}_{1000}^s$.
	         \label{figrninv}}
\end{figure}

In Fig. \ref{figrninv} we present the $\R^2$-embedded noisy time-series $x_{1:n}$, together
with the approximation of the stable set based on the noisy 
time-series $x_{1:n}$, via an HPDR color map. We remark  that the qualitative characteristics  
persist even when the approximation is based to short time-series contaminated by dynamical
noise.

\section{Discussion}\label{s5}

We have presented a novel approach for the stochastic approximation of the global stable manifold,
based on MCMC methods. Our approach is Bayesian, modeling the global stable manifold as the union
of supports of T-shifted to the past initial condition variables given appropriate sliding
window subsets of the time-series observations. 
By proper applications of the proposed SM-GSBR model, we were able to approximate the perturbed 
global stable manifold, and gain insight about the qualitative dynamical behavior in the zero 
noise limit, having as reference the associated deterministic system.

Our proposed SM-GSBR sampler is flexible because
it requites no prior knowledge on the type of the noise process perturbing the observed time-series. The reason for that, is that over the additive errors, 
we apply a Geometric Stick Breaking mixture process prior, which is supported over the space 
of symmetric zero-mean densities. In fact the SM-GSBR sampler requires no prior
information except for the model space required for the functional representation of the 
unknown deterministic part $g$ of the underlying map. Because we do not use any 
representation of the preimages of $g$ the SM-GSBR sampler is directly applicable to 
noninvertible maps.

A more versatile version of the SM-GSBR sampler, would involve relaxing the assumption 
of the functional representation of the unknown deterministic part $g$. For example by applying
over $g$ a Gaussian process \cite{rasmussen2004gaussian} prior.   
%\begin{enumerate}
%	\item Relaxing the assumption for a functional representation of the deterministic part, by adopting a Gaussian Process   prior supported over the space of functions. 
%	\item Improving manifold approximation, using manifold denoising algorithms \cite{hein2007manifold}, such as the Manifold Blurring Mean Shift algorithm \cite{wang2010manifold}.
	%	\item Extend the method in order to approximate the stable manifold in cases of more complex functional forms, where more sophisticated MCMC sampling schemes are required. 
%\end{enumerate}			

\appendix
\section{Proof of Proposition 1}\label{ap1}
To augment the posterior in (\ref{posterior1}), we will group the ${\cal V}_T$
auxiliary variables into $d_{-T+1:0}, N_{-T+1:0}$ and $d_{1:n}, N_{1:n}$.
We will also group the variables $\l,\tau_{1:\infty}$ and $\vt$, into the variable
$\nu=(\l,\tau_{1:\infty},\vt)$. The variables $x_{-T+1:0},\,d_{-T+1:0}$ and $N_{-T+1:0}$
to the variable $y_{-T+1:0}=(x_{-T+1:0},d_{-T+1:0},N_{-T+1:0})$, and the variables
$x_{1:n},\,d_{1:n}$ and $N_{1:n}$ to the variable $y_{1:n}=(x_{1:n},d_{1:n},N_{1:n})$.
%and admitting a prior distribution
%$(\l,\tau^\infty,\vt,\,x_{-T-d+1:-T})$
To truncate the intermediate variables between the $T$-shifted to 
the past initial condition and the observed time-series, we
condition the posterior density on the event ${\cal R}_T$.
Then, using Bayes' theorem, the posterior becomes
\begin{align}
&   \Pi(\nu,x_{-T-d+1:-T},y_{-T+1:0},d_{1:n},N_{1:n},\,|\,x_{1:n},{\cal R}_T)\,
     \propto\,\Pi(\nu,x_{-T-d+1:-T},y_{-T+1:0},y_{1:n}\,|\,{\cal R}_T)=\nonumber\\    
&   \Pi(\nu)\,\Pi(x_{-T-d+1:-T})\,\Pi(y_{-T+1:0}\,|\,
     \nu,\,x_{-T-d+1:-T},{\cal R}_T)\,\Pi(y_{1:n}\,|\,\nu,x_{-T-d+1:0})\propto\nonumber\\
&   \Pi(\nu)\prod_{1\le i\le d}{\cal I}(x_{-T-d+i}\in{\cal X})
    \left\{\Pi(y_{-T+1:0}\,|\,\nu,\,x_{-T-d+1:-T})
    \prod_{1\le i\le T}{\cal I}(x_{-T+i}\in{\cal X})\right\}
    \Pi(y_{1:n}\,|\,\nu,x_{-T-d+1:0})=\nonumber\\
&   \Pi(\nu)\prod_{1\le i\le T+d}{\cal I}(x_{-T-d+i}\in{\cal X})\,\,
    \Pi(y_{-T+1:n}\,|\,\nu,\,x_{-T-d+1:-T})\nonumber\\  
&   \Pi(\nu)\prod_{1\le i\le T+d}{\cal I}(x_{-T-d+i}\in{\cal X})\,\,
    \prod_{1\le i\le n+T}\Pi(x_{-T+i},d_{-T+i},N_{-T+i}\,|\,
    \nu,x_{-T+i-d:-T+i-1}).\nonumber           
\end{align}
Finally, the desired result comes from equation (\ref{xidiNi}). 
\hfill$\square$
%The variables $\nu$ and $x_{-T-d+1:-T}$ both admit prior distributions. 

% Create the reference section using BibTeX:
\bibliographystyle{plain}
\bibliography{bgsbr_arxiv}

\end{document}